\documentclass[prl,floatfix,superscriptaddress,amsmath,amssymb,
twocolumn,nobalancelastpage]{revtex4}

\usepackage{graphicx}
\usepackage{dcolumn}
\usepackage{bm}
\usepackage{pstricks}
\usepackage{datetime}

\newcommand{\ba}{\begin{eqnarray}}
\newcommand{\ea}{\end{eqnarray}}
\def\be{\begin{equation}}
\def\ee{\end{equation}}

\def\bimno{Bi$_3$Mn$_4$O$_{12}$(NO$_3$) }

\begin{document}

\title{Nematic quantum phases in the bilayer honeycomb antiferromagnet}

\author{Hao Zhang}
\email{zhanghao@iphy.ac.cn}
\affiliation{State Key Laboratory of Optoelectronic Materials and Technologies, School of Physics, Sun Yat-Sen University, Guangzhou 510275, China}

\author{ C. A.\ Lamas}
\email{lamas@fisica.unlp.edu.ar}
\affiliation{IFLP - CONICET, Departamento de F\'isica, Universidad Nacional de La Plata,
C.C.\ 67, 1900 La Plata, Argentina.}

\author{ M.\ Arlego}
\affiliation{IFLP - CONICET, Departamento de F\'isica, Universidad Nacional de La Plata,
C.C.\ 67, 1900 La Plata, Argentina.}

\author{ Wolfram Brenig}
\email{w.brenig@tu-bs.de}
\affiliation{Institute for Theoretical Physics, Technical University
Braunschweig, D-38106 Braunschweig, Germany}

\pacs{75.10.Jm, 75.50.Ee, 75.10.Kt}
\begin{abstract}
The spin$-1/2$ Heisenberg antiferromagnet on the honeycomb bilayer lattice is shown
to display a rich variety of semiclassical and genuinely quantum phases, controlled
by the interplay between intralayer frustration and interlayer exchange. Employing a
complementary set of techniques, comprising spin rotationally invariant Schwinger
boson mean field theory, bond operators, and series expansions we unveil the quantum
phase diagram, analyzing low-energy excitations and order parameters. By virtue of
Schwinger bosons we scan the complete range of exchange parameters, covering both
long range ordered as well as quantum disordered ground states and reveal the
existence of an extended, frustration induced lattice nematic phase in a range of
intermediate exchange unexplored so far.
\end{abstract}

\maketitle

Frustrated magnets are of great interest to a broad range of sub fields in physics,
harboring new quantum states of condensed matter \cite{Moessner2006,Balents2010},
fueling progress on fundamental paradigms of topological ordering
\cite{Wen1990,Wen2002,Kitaev2003}, providing realistic prospects for quantum
computing \cite{Benjamin2015,Riste2015,Corcoles2015} and devices for thermal
management technologies \cite{Zhitomirsky2003,Pakhira2017}, inspiring research on
ultra cold atomic gases \cite{Kim2010,Struck2011}, realizing elementary excitations
related to Grand Unified Theories (GUT) \cite{Castelnovo2008,Mengotti2011,Hooft1974}, and exhibiting
correlations found in soft matter, liquid crystals, and even cosmic strings
\cite{Jaubert2011,Austin1994}.  Strong frustration in quantum magnets can ultimately
lead to spin liquids, free of any broken symmetries, featuring long-range
entanglement, topological order and anyonic excitations
\cite{Balents2010,Savary2016}.  Proximate to such liquids, a rich variety of
additional exotic quantum matter, including valence bond crystals, also termed {\it
lattice nematics} \cite{Matan2010}, chiral liquids \cite{Grohol2005}, multipolar
states \cite{Damle2006}, and more complex phases compete for stability.
Understanding such phases of matter and their interplay is a critically outstanding
problem for theory and experiment. In this letter we take a major step forward into
this direction and detail the emergence of lattice nematic order in a yet unexplored
region of frustrated magnets on bilayer honeycomb lattices.

Recently, frustrated Heisenberg models on single layer honeycomb lattices have become
a test-bed for competing spiral order, lattice nematicity and plaquette valence bond
states \cite{Mulder,Okumura,Wang,Mosadeq,Cabra_honeycomb_prb, Ganesh_2011,
Albuquerque,Clark, Cabra_honeycomb_2, Mezzacapo,Bishop_2012,Li_2012_honeyJ1-J2-J3,
Bishop_2013,Fisher_2013, Ganesh_PRL_2013,Zhu_PRL_2013,Zhang_PRB_2013}.  This interest
has been propelled by the discovery of bismuth oxynitrate, {\bimno}
\cite{smirnova2009synthesis}, where Mn$^{4+}$ ions of spin $3/2$ form honeycomb
layers, with both, nearest and next-nearest neighbor antiferromagnetic (AFM)
exchange. Early on however, it was noticed that in this compound Mn$^{4+}$ ions are
grouped into pairs along the c-axis, rendering the structure rather that of a bilayer
honeycomb lattice. Despite a significant separation through bismuth atoms, density
functional calculation \cite{kandpal2011calculation} resulted in comparable inter-
and intralayer exchange, consistent with experimental findings \cite{expnew2}. This
has lead to first investigations of bilayer honeycomb systems \cite{Ganesh_QMC,
Oitmaa_2012, Zhang2014, Arlego201415, Brenig2016, bishop2017frustrated,
Richter2017}. Most of these studies have focused on the stability of the
semi-classical phases, extending previous work on the single layer case.

First indications of quantum disordered phases, genuinely related to the bilayer
geometry and not present in the single layer case have been provided in a small
parameter window in \cite{Brenig2016}, following ideas of \cite{Bose1992,Bose1993a}
and similar works \cite{Honecker2000a, Brenig2001a, Arlego2006a, Lamas2015b,matera-lamas2}. However
a complete understanding of the quantum phase diagram of the bilayer is
missing. Therefore, in this letter we provide a comprehensive analysis of the quantum
phases of the frustrated Heisenberg model on the honeycomb bilayer over a wide range
of coupling strengths, including in particular the intermediate regime, where both,
the interlayer exchange and intralayer frustration are comparable to the intralayer
first neighbor couplings. This part of the phase diagram has remained largely
unexplored, representing a challenge for most of the existing state-of-the-art numerical
techniques.  Here, by means of a combination of methods, among which Schwinger Bosons
stand out for their ability to explore the full quantum phase diagram and to treat on
equal footing quantum and semi-classical states, we unveil a rich structure of
phases, where the interplay of frustration and interlayer coupling is most essential,
destroying magnetic order and giving rise to exotic phases. Most noteworthy, we
will provide evidence for a new lattice nematic phase in the intermediate coupling
regime.

\begin{center}
\begin{figure}[t]
\includegraphics[width=0.9\columnwidth]{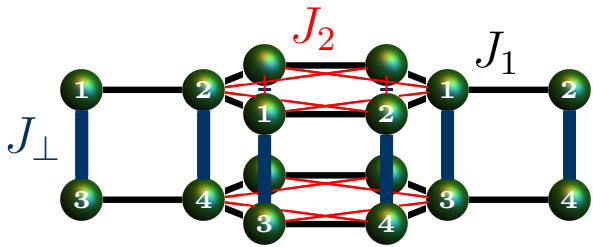}
\caption{(Color online)  Schematic representation of the model.
The sites (green spheres) in each unit cell are labeled from 1 to 4. Bold vertical blue lines indicate $J_{\bot}$ interlayer couplings,
whereas thin black and red lines indicate $J_1$ and $J_2$ first and second nearest neighbors, respectively. }
\label{fig:bilayer}
\end{figure}
\end{center}

The model we consider is shown in Fig. \ref{fig:bilayer}. Its Hamiltonian reads
\ba
\label{eq:Hamiltonian}
H=\sum_{i,l,m}J^{(l,m)}_{i}\vec{\bf{S}}_{l}(\vec{r})\cdot \vec{\bf{S}}_{m}(\vec{r}+\vec{e}_{i}),
\ea
$\vec{r}$ and $\vec{e}_{i}$ label sites and primitive vectors of the triangular
Bravais lattice. $\vec{\bf{S}}_{l}(\vec{r})$ are spin operators at basis sites
$\vec{r}$, $l{=}1...4$ of the bilayer. The couplings $J^{(l,m)}_{i}$ are non-zero,
with values $J_{\bot}$, $J_1$ and $J_2$ as depicted in Fig. \ref{fig:bilayer}.

%

\begin{figure}
\includegraphics[width=0.85\columnwidth]{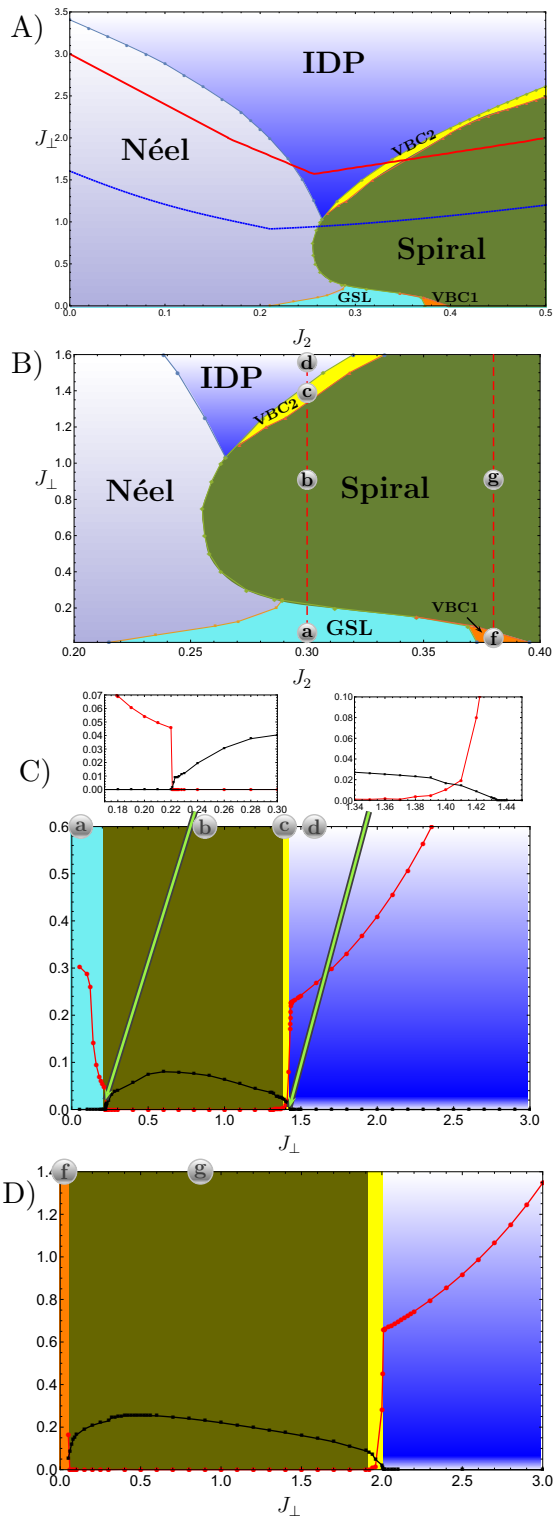}
\caption{(Color online) Panel (A): the different colored regions identify the
phases of model (1) in $J_{2}-J_{\bot}$ plane (in units of $J_1$) determined by
SBMF.  The blue and red lines represent the border of the IDP phase
predicted by SE and BO respectively.  Panel (B): Zoom of the phase diagram where
two paths along $J_{\bot}$, for $J_2 = 0.3$ and $J_2 = 0.38$ are indicated with
red dashed lines.  Lower panels (C-D), depict the evolution of the gap
(connected red dots) and the $ Z_3 $ directional symmetry-breaking order
parameter $\rho$ (connected black dots), along the mentioned paths.}
\label{fig:Phasediag_comp}
\end{figure}

Before describing our calculations, we focus on the main results, summarized in
Fig. \ref{fig:Phasediag_comp}A. On the classical level, $S \rightarrow \infty $, and
in the single plane limit, i.e. at $J_{\bot} = 0 $, there are two phases. N\'eel
order for $ J_2 / J_1 < 1/6 $ and spiral order for $ J_2 / J_1 > 1/6 $. Allowing for
interlayer coupling this single transition point extends into a line, {\it
independent} of $J_{\bot}$. Quantum fluctuations lead to new non-classical
intermediate phases and renormalize the N\'eel and spiral phases.  Previous studies
\cite{Zhang_PRB_2013} have identified continuous transitions into two frustration
induced genuine quantum phases. A gapped spin liquid (GSL) phase, preserving all
lattice symmetries for $0.2075 \lesssim J_2 / J_1 \lesssim 0.3732$ and a
staggered-dimer lattice nematic phase (VBC1) which maintains the SU(2) spin rotational
and lattice translational symmetries, but breaks $Z_3$ symmetry, corresponding to $2
\pi/3$ rotations around an axis perpendicular to the plane for $0.3732 \lesssim J_2 /
J_1 \lesssim 0.398$.  Another limiting case is $J_{\bot} \rightarrow \infty$. Here an
interlayer dimer product phase (IDP) is formed.

Connecting these two limits, the interplay between the interlayer couplings
$J_{\bot}$ and the frustration $J_2$ reveals the complex phase diagram we find
in Fig. \ref{fig:Phasediag_comp}A. Starting from the limit of decoupled planes,
we consider the semiclassical N\'eel and spiral phases first.  Figure
\ref{fig:Phasediag_comp}A shows, that small interlayer couplings extend each of
their windows of stability along the $J_2$ direction, leading even to a region
of competition. However for sufficiently large $J_{\bot}$ the semiclassical
phases recess and are suppressed in favor of the IDP. Regarding the GSL and VBC1
phases, interlayer coupling has a dramatic consequence, suppressing them very
rapidly, reentering semiclassical phases at finite $J_{\bot}$.  Finally, the
N\'eel-to-IDP is direct. This is not so for the Spiral-to-IPD transition. In
fact we find yet another lattice nematic region (VBC2) which intervenes. To the
best of our knowledge, this has not been observed before.

Next we detail how to arrive at our main result, i.e. Fig. \ref{fig:Phasediag_comp},
using three complementary techniques, namely, Schwinger bosons mean field theory
(SBMFT)\cite{Cabra_honeycomb_prb}, Bond operators (BO)\cite{Sachdev1990} and series
expansions (SE) \cite{SE-CUT}. While being a mean field approach, primarily gauged
towards bosonic fixed point models, SBMFT is superior in addressing on equal footing
both semi-classical as well as genuinely quantum phases, allowing to scan all of
Eq. (\ref{eq:Hamiltonian})'s parameter space. The other two methods are best suited
to obtain additional information, for large interlayer coupling (BO), and for weak
frustration (SE).

\noindent {\it SBMFT.---} Here, spin operators are represented by two bosons
$\vec{\mathbf{S}}_{l}(\vec{r})=$ $\frac{1}{2}\vec{\mathbf{b}}_{l}^{\dag}(\vec{r})
\cdot\vec{\sigma}\cdot\vec{\mathbf{b}}_{l}(\vec{r})$ \cite{Auerbach, Auerbach:1994,
Auerbach:2011}, where ${\vec{\bf b}^{\dagger }_{l}(\vec{r})} \!= \!({\bf b}^{\dagger
}_{l,\uparrow }(\vec{r}),{\bf b}^{\dagger }_{l,\downarrow }(\vec{r}))$ is a spinor,
$\vec{\sigma}$ are the Pauli matrices, and $\sum_\sigma \mathbf{b}^{\dag}_{l,\sigma}
(\vec{r}) \mathbf{b}_{l,\sigma}(\vec{r})\!=\!2S$ is a local constraint.  Using the
rotationally invariant representation \cite{Zhang_PRB_2013, Cabra_honeycomb_prb,
Cabra_honeycomb_2, Trumper1, Trumper2, Coleman, Mezio, Messio, Messio_2013}, we
define two $SU(2)$ invariants $\mathbf{A}_{l m}(\vec{x},\vec{y})=\frac12
\sum_{\sigma} \sigma \mathbf{b}_{l,\sigma}(\vec{x})\mathbf{b}_{m,-\sigma}(\vec{y})$
and $\mathbf{B}_{l m}(\vec{x},\vec{y})=\frac12 \sum_{\sigma}
\mathbf{b}^{\dag}_{l,\sigma} (\vec{x})\mathbf{b}_{m,-\sigma} (\vec{y})$, where the
former generates a spin singlet between sites $l$ and $m$, and the latter a coherent
hopping of the Schwinger bosons.
At the mean field level, the exchange follows as $\langle
(\vec{\mathbf{S}}_{l}(\vec{x})\cdot \vec{\mathbf{S}}_{m}(\vec{y}) )_{MF} \rangle =
|B_{l m}(\vec{x}-\vec{y})|^{2}-|A_{l m}(\vec{x}-\vec{y})|^{2},$ with $A^{*}_{l
m}(\vec{x}-\vec{y})=\langle \mathbf{A}^{\dag}_{l m}(\vec{x},\vec{y})\rangle$ and
$B^{*}_{l m}(\vec{x}-\vec{y})=\langle \mathbf{B}^{\dag}_{l
m}(\vec{x},\vec{y})\rangle$.
These equations are solved self-consistently taking into account the constraint in
the number of bosons $B_{ll}(\vec{R}=\vec{0})=4N_{c}S$, whith $N_c$ representing the
total number of unit cells and $S$ the spin strength
\cite{Cabra_honeycomb_prb,Zhang_PRB_2013}.

After solving the mean-field equations on finite but large lattices we primarily
extract the extrapolation of the elementary excitation gap $\Delta$. This is
used to classify magnetic phases, for which $\Delta$ has to be zero. If
$\Delta\neq 0$, Bose condensation cannot occur and the phase is quantum
disordered. We can also obtain the real space spin correlation function $C_l$
and magnetization $m_l$ \cite{note:corr}. Lattice nematic phases, which preserve
the lattice translational invariance, but break $Z_3$ lattice symmetry have come
under scrutiny early on in single layer honeycomb
systems. These imply a nonzero order parameter $\rho=\frac43 |(\langle
\vec{\mathbf{S}}_{1}(\vec{r}) \cdot \vec{\mathbf{S}}_{2}(\vec{r}) \rangle
+e^{i2\pi/3} \langle \vec{\mathbf{S}}_{1}(\vec{r}) \cdot
\vec{\mathbf{S}}_{2}(\vec{r}+\vec{e}_{1}) \rangle + e^{i4\pi/3} \langle
\vec{\mathbf{S}}_{1}(\vec{r}) \cdot \vec{\mathbf{S}}_{2}(\vec{r}-\vec{e}_{2}) \rangle
)|$ \cite{Okumura,Mulder}. Here we have investigated this order parameter over {\it
all} of the parameter space of Fig. \ref{fig:Phasediag_comp}A.

To clarify the procedure we detail our SBMFT results along the two paths (a-d),
at $J_2 = 0.3$ and (f-g), at $J_2 = 0.38$ in a representative part of the phase
diagram, depicted in Fig. \ref{fig:Phasediag_comp}B. The corresponding evolution
of $\Delta$ (connected red dots) and $\rho$ (connected black dots) are shown in
Fig. \ref{fig:Phasediag_comp}C and \ref{fig:Phasediag_comp}D for the paths (a-d)
and (f-g), respectively.

We start in the blue-light phase around point (a), where panel C) features a
finite gap and unbroken $Z_3$ symmetry. This identifies the GSL phase. As
$J_{\bot}$ increases from zero, the gap rapidly decreases and closes {\it
simultaneously} with $\rho$ growing finite. A blowup of this is shown in the
left upper inset of Fig. \ref{fig:Phasediag_comp}C. This behavior is consistent
with a spiral phase, which is gapless and breaks $Z_3$ symmetry. As $J_{\bot}$
increases further, $\rho$ runs through a maximum and decreases up to a point
where $\Delta\neq 0$ again. In stark contrast to the GSL however a narrow yellow
region of broken $Z_3$ symmetry {\it and} gapful behavior surfaces around point
(c).   This characterizes the lattice nematic phase (VBC2).  Right upper inset
of Fig. \ref{fig:Phasediag_comp}C shows a blowup of this region. Very different
than the VBC1 phase, VBC2 can be found in a much larger range of parameters
running all along the upper spiral phase boundary. Finally, entering the blue
region around point (d), there is a near 1st order jump to rather large values
of $\Delta$ where $\rho$ turns zero, restoring $Z_3$ symmetry. This is
consistent with the IDP, adiabatically connecting to the limit of decoupled
dimers at $J_{\bot}=\infty$

Turning to the second path (points f and g), it is obvious from
Fig. \ref{fig:Phasediag_comp}D, that the phases corresponding to point (g) is
identical with the corresponding one at point (b). However, different from the GSL at
(a), the phase VBC1 around (f) displays a behavior of $\Delta$ and $\rho$ identical
to point (c), i.e. a lattice nematic. We cannot exclude the existence of observables
beyond our study which allow for further discrimination between VBC1 and VBC2.

\begin{center}
\begin{figure*}[tb]
\includegraphics[width=1.9\columnwidth]{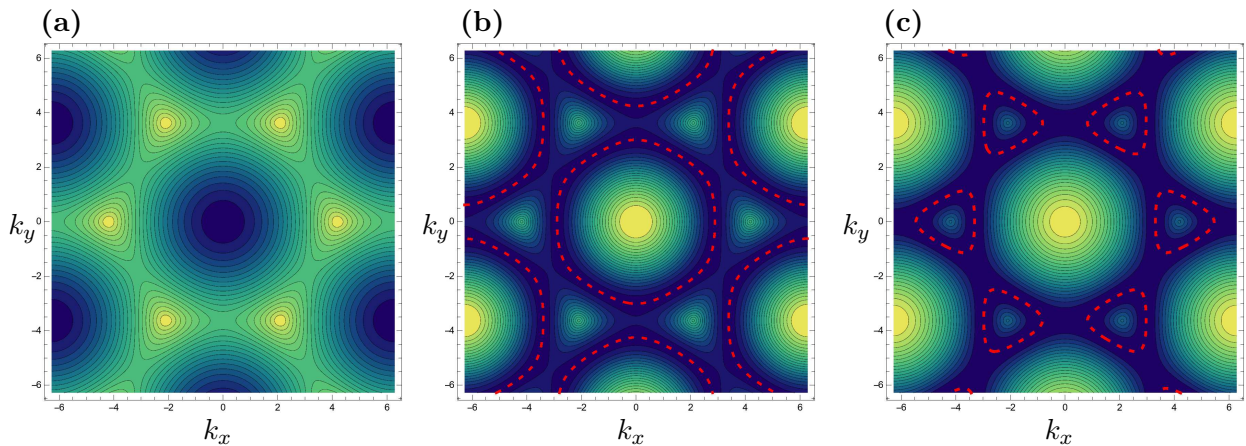}
\caption{(Color online) Contour plot corresponding to the Bond operator boson
dispersion close to condensation for (a): $J_2/J_{1}=0.1$, (b) $J_2/J_{1}=0.4$ and
(c) $J_2/J_{1}=0.6$. Red dashed lines correspond to curves in the k space determining
the classical manifold of spiral ground states.} \label{fig:contour-BO}
\end{figure*}
\end{center}

\noindent {\it Series Expansion and Bond Operator Approach.---} For a complementary
analysis of the evolution of the quantum disordered phases, starting from the limit of
decoupled dimers, $J_{\bot}\rightarrow \infty$, we use both, series expansion (SE)
\cite{Book-SE} and bond operator theory (BOT)
\cite{Chubukov1989,Chubukov1991,Sachdev1990}.  In BOT, spins at the vertices of each
dimer are written as $S^{\alpha}=(\pm s^{\dagger}t_{\alpha}\pm
t^{\dagger}s-\sum_{\beta,\gamma}i \varepsilon_{ \alpha\beta\gamma}
t_{\beta}^{\dagger}t_{\gamma}^{\phantom{\dagger}})/2$, with the constraint
$s^{\dagger}s + \sum_{\alpha} t_{\alpha}^{\dagger} t_{\alpha}^{
\phantom{\dagger}}=1$, where $s^{\dagger}(t_{\alpha}^{\dagger})$ create
singlet(triplet) states of the dimer, and $\alpha=1,2,3$ labels the triplet multiplet.
BOT maps the spin model onto an interacting Bose gas, for which several schemes of
treatment have been proposed \cite{Sachdev1990,Chubukov1989,Chubukov1991,Kotov1998,Matsumoto2002}. Here we use the Holstein-Primakoff (HP)
approximation \cite{Chubukov1989,Chubukov1991}, where $s^{(\dagger)}$ is replaced by
a $C$-number and $s =(1 - \sum_{\alpha} t_{\alpha}^{\dagger} t_{\alpha}^{
\phantom{\dagger}})^{1/2}$ is expanded to obtain a quadratic triplon
Hamiltonian. Standard Bogoliubov diagonalization yields a ground state energy per unit cell of
$E={-}\frac{9}{4}+\frac{3}{4N}\sum_{k,\pm} [ 1 \pm \epsilon_{\pm}( k ) ]^{1/2}$, with the
triplon dispersions $\epsilon_{\pm}(k)=$ $\frac{J_1}{J_{\bot}} [3{+}$ $2\cos(k_{x})+
4\cos(k_{x}/2)$ $\cos(\sqrt{3}k_{y}/2)]^{1/2}$ $\pm 2 \frac{J_1}{J_{\bot}}
(\cos(k_{x})+$ $2\cos(k_{x}/2)$ $\cos(\sqrt{3}k_{y}/2))$. At $J_1 {=}J_2{=}0$, and for
the latter one recovers the bare singlet-triplet gap $\Delta=J_{\bot}$, and for
the former $E={-}3J_{\bot}/4$ consistent with a bare singlet.

For SE we use the continuous unitary transformation (CUT) method \cite{Wegner-1994a,
SE-CUT, Arlego2011, Arlego2008a, Arlego2007a, Arlego2006a} starting from the limit of
decoupled dimers. This method allows to obtain analytical expressions for the ground
state energy and the dispersion of the elementary triplon excitations of the IDP
versus $J_{1,2}/J_{\bot}$. We have evaluated these up to $O(4)$. Their rather lengthy
expressions are detailed elsewhere \cite{SuppMat}.

In Fig. \ref{fig:Phasediag_comp}A we show the critical lines of the gap closure of
the triplon dispersion of the IDP obtained from both, BOT-HP (red solid) and SE-CUT (blue dotted line). Clearly,
contrasting them with regions of magnetic ordering obtained from SBMFT, the general
trend, i.e. the breakdown of magnetic order versus $J_2$ and $J_{\bot}$ is fully
consistent with the IDP gap closure. Quantitatively however, comparing SBMFT and
BOT-HP to SE-CUT, the latter predicts a smaller range of stability for semiclassical
phases. Since the former two are mean-field theories, such a tendency to prefer
ordered phases is a well known shortcoming. In fact, SE-CUT locates the IDP-N\'eel
transition at $J_{\bot} \simeq 1.6$ for $J_2 = 0$ in Fig. \ref{fig:Phasediag_comp}A,
in excellent agreement with quantum Monte-Carlo calculations \cite{Ganesh_QMC}, and
moreover SE-CUT is rather close to coupled-cluster results for finite, but small
$J_2\lesssim 0.2$ \cite{bishop2017frustrated}. For larger $J_2$ the CUT-SE
bcomes less reliable and we remain with only BOT-HP to compare to within the IDP.

Finally, we comment on the location in $\vec{k}$-space of Bose condensation within
the BOT-HP, as compared to the classical magnetic pitch vector $\vec{Q}$ of the
bilayer at $S\rightarrow \infty$. As mentioned previously, the latter is independent
of $J_{\bot}$, comprising a N\'eel state for each plane for $0< J_{2}/J_{1} <1/6$ and
for $1/6< J_{2}/J_{1}$ a set of classically degenerate coplanar spiral ground states
with $\vec{S}_l(\vec{r})=(-1)^l S[ \cos(\vec{Q}\cdot\vec{r}+\theta_l)\hat{i}
+\sin(\vec{Q}\cdot\vec{r}+\theta_l)\hat{j}]$, where the pitch vector lies on the
closed curve $\cos(Q_x)+\cos(-Q_x/2+\sqrt{3}Q_y/2)+\cos(Q_x/2+ \sqrt{3}Q_y/2)+3/2=
(J_1/J_2)^2/8$, and the phase $\theta_i$ obeys $\theta_{1,2}=\pi+\theta_{3,4}$
\cite{Mulder}. Comparing this now, to the critical wave vector $\vec{Q}$ for Bose
condensation within the BOT-HP, we first have $\vec{Q}=(0,0)$ corresponding to a
N\'eel order for $J_{2}<\frac{1}{6}$. For $J_{2}>\frac{1}{6}$, condensation does not
occur at a single point, but on lines in $\vec{k}$-space. Remarkably, these are
identical to those from the classical states.  This is illustrated in
Fig. \ref{fig:contour-BO}(a-c), where we plot contours of the boson dispersion close
to condensation at $J_{2}=0.1,0.4$ and $0.6$ and incorporate the degenerate
classical spiral pitch vector locations by red dashed lines. In turn, while quantum
fluctuations may modify such agreement, it is nevertheless interesting to realize
that BOT-HP provides some guidance as to the type of semiclassical phase to emerge
upon gap closure.

In conclusion, the interplay between intralayer frustration and interlayer exchange
allows for a rich variety of classical and quantum disordered phases to compete for
stability in the bilayer honeycomb antiferromagnet. In this letter, evidence for
these phases has been provided over a wide range of coupling constants using three
complementary methods, yielding consistent results. Most noteworthy, at intermediate
coupling, we have discovered a new lattice nematic phase, which exists in a region of
parameter space substantially larger than similar phases observed previously in the model
at small interlayer coupling. While we have carried out our analysis on a
spin-$1/2$ model, our findings may be relevant to understand the absence of magnetic
order in the spin-$3/2$ honeycomb bilayer materials \bimno , where first principle
calculations suggest exchange paths identical to our microscopic model and to be all
of similar magnitude.

\emph{Acknowledgments: }
C.A.L gratefully acknowledges the support of NVIDIA Corporation.
C.A.L. and M.A. are supported by CONICET (PIP 1691) and ANPCyT (PICT
2013-0009). H.Z. thanks Lu Yu for fruitful discussions, and the Institute of Physics,
Chinese Academy of Sciences for financial support. Work of W.B. has been supported in
part by the DFG through SFB 1143. W.B. also acknowledges kind hospitality of the
PSM, Dresden.


\begin{thebibliography}{99}
\expandafter\ifx\csname natexlab\endcsname\relax\def\natexlab#1{#1}\fi
\expandafter\ifx\csname bibnamefont\endcsname\relax
  \def\bibnamefont#1{#1}\fi
\expandafter\ifx\csname bibfnamefont\endcsname\relax
  \def\bibfnamefont#1{#1}\fi
\expandafter\ifx\csname citenamefont\endcsname\relax
  \def\citenamefont#1{#1}\fi
\expandafter\ifx\csname url\endcsname\relax
  \def\url#1{\texttt{#1}}\fi
\expandafter\ifx\csname urlprefix\endcsname\relax\def\urlprefix{URL }\fi
\providecommand{\bibinfo}[2]{#2}
\providecommand{\eprint}[2][]{\url{#2}}

\bibitem{Moessner2006}
R. Moessner and A.P. Ramirez,
Phys. Today {\bf 59}, 24 (2006).

\bibitem{Balents2010}
L. Balents, Nature {\bf 464}, 199 (2010).

\bibitem{Wen1990}
X.-G. Wen,
Int. J. Mod. Phys. B, {\bf 04}, 239 (1990).

\bibitem{Wen2002}
X.-G. Wen,
Phys. Rev. B {\bf 65}, 165113 (2002).

\bibitem{Kitaev2003}
A. Kitaev, Ann. Phys. (N.Y.) {\bf 303}, 2 (2003).

\bibitem{Benjamin2015}
S. Benjamin and J. Kelly,
Nat. Mater. {\bf 14}, 561 (2015).

\bibitem{Riste2015}
D. Rist\`e, S. Poletto, M.-Z. Huang, A. Bruno, V. Vesterinen,
O.-P. Saira, and L. DiCarlo, Nat. Commun. {\bf 6}, 7983 (2015).

\bibitem{Corcoles2015}
A. D. C\'orcoles, E. Magesan, S. J. Srinivasan, A. W. Cross, M. Steffen,
J. M. Gambetta, and J. M. Chow, Nat. Commun. {\bf 6}, 7979 (2015).


\bibitem{Zhitomirsky2003}
M. E. Zhitomirsky,
Phys. Rev. B {\bf 67}, 104421 (2003).

\bibitem{Pakhira2017}
S. Pakhira, C. Mazumdar, R. Ranganathan, and M. Avdeev,
Nature Sci. Rep. {\bf 7}, 7367 (2017).

\bibitem{Kim2010}
K. Kim, M.-S. Chang, S. Korenblit, R. Islam, E. E. Edwards,
J. K. Freericks, G.-D. Lin, L.-M. Duan, and C. Monroe,
Nature {\bf 465}, 590 (2010).

\bibitem{Struck2011}
J. Struck, C. Ölschläger, R. L. Targat, P. Soltan-Panahi,
A. Eckardt, M. Lewenstein, P. Windpassinger, and K. Sengstock,
Science {\bf 333}, 996 (2011).


\bibitem{Hooft1974}
G. 't Hooft,
Nucl. Phys. B {\bf 79} 276 (1974).

\bibitem{Castelnovo2008}
C. Castelnovo, R. Moessner, and S. L. Sondhi,
Nature {\bf 451}, 42 (2008)

\bibitem{Mengotti2011}
E. Mengotti, L. J. Heyderman, A. F. Rodr\'iguez, F. Nolting, R. V. H\"ugli, and
H.-B. Braun, Nat. Phys. {\bf 7} 68 (2011).

\bibitem{Jaubert2011}
L. D. C. Jaubert, M. Haque, and R. Moessner,
Phys. Rev. Lett. {\bf 107}, 177202 (2011).

\bibitem{Austin1994}
D. Austin, E. J. Copeland, and R. J. Rivers,
Phys. Rev. D {\bf 49}, 4089 (1994).

\bibitem{Savary2016}
L. Savary and L. Balents,
Rep. Prog. Phys. {\bf 80}, 016502 (2017).



\bibitem{Matan2010}
K. Matan, T. Ono, Y. Fukumoto, T. J. Sato, J. Yamaura, M. Yano,
K. Morita, and H. Tanaka,
Nat. Phys. {\bf 6}, 865 (2010).

\bibitem{Grohol2005}
D. Grohol, K. Matan, J.-H. Cho, S.-H. Lee, J. W. Lynn, D. G. Nocera,
and Y. S. Lee,
Nat. Mater. {\bf 4}, 323 (2005).

\bibitem{Damle2006}
K. Damle and T. Senthil,
Phys. Rev. Lett. {\bf 97}, 067202 (2006).




\bibitem[{\citenamefont{Mulder et~al.}(2010)\citenamefont{Mulder, Ganesh,
  Capriotti, and Paramekanti}}]{Mulder}
\bibinfo{author}{\bibfnamefont{A.}~\bibnamefont{Mulder}},
  \bibinfo{author}{\bibfnamefont{R.}~\bibnamefont{Ganesh}},
  \bibinfo{author}{\bibfnamefont{L.}~\bibnamefont{Capriotti}},
  \bibnamefont{and}
  \bibinfo{author}{\bibfnamefont{A.}~\bibnamefont{Paramekanti}},
  \bibinfo{journal}{Phys. Rev. B} \textbf{\bibinfo{volume}{81}},
  \bibinfo{pages}{214419} (\bibinfo{year}{2010}).

\bibitem[{\citenamefont{Okumura et~al.}(2010)\citenamefont{Okumura, Kawamura,
  Okubo, and Motome}}]{Okumura}
\bibinfo{author}{\bibfnamefont{S.}~\bibnamefont{Okumura}},
  \bibinfo{author}{\bibfnamefont{H.}~\bibnamefont{Kawamura}},
  \bibinfo{author}{\bibfnamefont{T.}~\bibnamefont{Okubo}}, \bibnamefont{and}
  \bibinfo{author}{\bibfnamefont{Y.}~\bibnamefont{Motome}},
  \bibinfo{journal}{J.\ Phys.\ Soc.\ Jpn.} \textbf{\bibinfo{volume}{79}},
  \bibinfo{pages}{114705} (\bibinfo{year}{2010}).

\bibitem[{\citenamefont{Wang}(2010)}]{Wang}
\bibinfo{author}{\bibfnamefont{F.}~\bibnamefont{Wang}}, \bibinfo{journal}{Phys.
  Rev. B} \textbf{\bibinfo{volume}{82}}, \bibinfo{pages}{024419}
  (\bibinfo{year}{2010}).

\bibitem[{\citenamefont{Mosadeq et~al.}(2011)\citenamefont{Mosadeq, Shahbazi,
  and Jafari}}]{Mosadeq}
\bibinfo{author}{\bibfnamefont{H.}~\bibnamefont{Mosadeq}},
  \bibinfo{author}{\bibfnamefont{F.}~\bibnamefont{Shahbazi}}, \bibnamefont{and}
  \bibinfo{author}{\bibfnamefont{S.}~\bibnamefont{Jafari}},
  \bibinfo{journal}{Journal of Physics: Condensed Matter}
  \textbf{\bibinfo{volume}{23}}, \bibinfo{pages}{226006}
  (\bibinfo{year}{2011}).

\bibitem[{\citenamefont{Cabra et~al.}(2011{\natexlab{a}})\citenamefont{Cabra,
  Lamas, and Rosales}}]{Cabra_honeycomb_prb}
\bibinfo{author}{\bibfnamefont{D.~C.} \bibnamefont{Cabra}},
  \bibinfo{author}{\bibfnamefont{C.~A.} \bibnamefont{Lamas}}, \bibnamefont{and}
  \bibinfo{author}{\bibfnamefont{H.~D.} \bibnamefont{Rosales}},
  \bibinfo{journal}{Phys. Rev. B} \textbf{\bibinfo{volume}{83}},
  \bibinfo{pages}{094506} (\bibinfo{year}{2011}{\natexlab{a}}).

\bibitem[{\citenamefont{Ganesh et~al.}(2011{\natexlab{a}})\citenamefont{Ganesh,
  Sheng, Kim, and Paramekanti}}]{Ganesh_2011}
\bibinfo{author}{\bibfnamefont{R.}~\bibnamefont{Ganesh}},
  \bibinfo{author}{\bibfnamefont{D.}~\bibnamefont{Sheng}},
  \bibinfo{author}{\bibfnamefont{Y.-J.} \bibnamefont{Kim}}, \bibnamefont{and}
  \bibinfo{author}{\bibfnamefont{A.}~\bibnamefont{Paramekanti}},
  \bibinfo{journal}{Phys.\ Rev.\ B} \textbf{\bibinfo{volume}{83}},
  \bibinfo{pages}{144414} (\bibinfo{year}{2011}{\natexlab{a}}).

\bibitem[{\citenamefont{Albuquerque et~al.}(2011)\citenamefont{Albuquerque,
  Schwandt, Het\'{e}nyi, Capponi, Mambrini, and L\"auchli}}]{Albuquerque}
\bibinfo{author}{\bibfnamefont{A.~F.} \bibnamefont{Albuquerque}},
  \bibinfo{author}{\bibfnamefont{D.}~\bibnamefont{Schwandt}},
  \bibinfo{author}{\bibfnamefont{B.}~\bibnamefont{Het\'{e}nyi}},
  \bibinfo{author}{\bibfnamefont{S.}~\bibnamefont{Capponi}},
  \bibinfo{author}{\bibfnamefont{M.}~\bibnamefont{Mambrini}}, \bibnamefont{and}
  \bibinfo{author}{\bibfnamefont{A.~M.} \bibnamefont{L\"auchli}},
  \bibinfo{journal}{Phys. Rev. B} \textbf{\bibinfo{volume}{84}},
  \bibinfo{pages}{024406} (\bibinfo{year}{2011}).

\bibitem[{\citenamefont{Clark et~al.}(2011)\citenamefont{Clark, Abanin, and
  Sondhi}}]{Clark}
\bibinfo{author}{\bibfnamefont{B.}~\bibnamefont{Clark}},
  \bibinfo{author}{\bibfnamefont{D.}~\bibnamefont{Abanin}}, \bibnamefont{and}
  \bibinfo{author}{\bibfnamefont{S.}~\bibnamefont{Sondhi}},
  \bibinfo{journal}{Phys. Rev. Lett.} \textbf{\bibinfo{volume}{107}},
  \bibinfo{pages}{087204} (\bibinfo{year}{2011}).

\bibitem[{\citenamefont{Cabra et~al.}(2011{\natexlab{b}})\citenamefont{Cabra,
  Lamas, and Rosales}}]{Cabra_honeycomb_2}
\bibinfo{author}{\bibfnamefont{D.}~\bibnamefont{Cabra}},
  \bibinfo{author}{\bibfnamefont{C.}~\bibnamefont{Lamas}}, \bibnamefont{and}
  \bibinfo{author}{\bibfnamefont{H.}~\bibnamefont{Rosales}},
  \bibinfo{journal}{Mod. Phys. Lett. B} \textbf{\bibinfo{volume}{25}},
  \bibinfo{pages}{891} (\bibinfo{year}{2011}{\natexlab{b}}).

\bibitem[{\citenamefont{Mezzacapo and Boninsegni}(2012)}]{Mezzacapo}
\bibinfo{author}{\bibfnamefont{F.}~\bibnamefont{Mezzacapo}} \bibnamefont{and}
  \bibinfo{author}{\bibfnamefont{M.}~\bibnamefont{Boninsegni}},
  \bibinfo{journal}{Phys. Rev. B} \textbf{\bibinfo{volume}{85}},
  \bibinfo{pages}{060402(R)} (\bibinfo{year}{2012}).

\bibitem[{\citenamefont{Bishop et~al.}(2012)\citenamefont{Bishop, Li., Farnell,
  and Campbell}}]{Bishop_2012}
\bibinfo{author}{\bibfnamefont{R.~F.} \bibnamefont{Bishop}},
  \bibinfo{author}{\bibfnamefont{P.~H.~Y.} \bibnamefont{Li.}},
  \bibinfo{author}{\bibfnamefont{D.~J.~J.} \bibnamefont{Farnell}},
  \bibnamefont{and} \bibinfo{author}{\bibfnamefont{C.~E.}
  \bibnamefont{Campbell}}, \bibinfo{journal}{J.\ Phys.: Condens.\ Matter}
  \textbf{\bibinfo{volume}{24}}, \bibinfo{pages}{236002}
  (\bibinfo{year}{2012}).

\bibitem[{\citenamefont{Li et~al.}(2012)\citenamefont{Li, Bishop, Farnell, and
  Campbell}}]{Li_2012_honeyJ1-J2-J3}
\bibinfo{author}{\bibfnamefont{P.~H.~Y.} \bibnamefont{Li}},
  \bibinfo{author}{\bibfnamefont{R.~F.} \bibnamefont{Bishop}},
  \bibinfo{author}{\bibfnamefont{D.~J.~J.} \bibnamefont{Farnell}},
  \bibnamefont{and} \bibinfo{author}{\bibfnamefont{C.~E.}
  \bibnamefont{Campbell}}, \bibinfo{journal}{Phys.\ Rev.\ B}
  \textbf{\bibinfo{volume}{86}}, \bibinfo{pages}{144404}
  (\bibinfo{year}{2012}).

\bibitem[{\citenamefont{Bishop et~al.}(2013)\citenamefont{Bishop, Li, Farnell,
  and Campbell}}]{Bishop_2013}
\bibinfo{author}{\bibfnamefont{R.~F.} \bibnamefont{Bishop}},
  \bibinfo{author}{\bibfnamefont{P.~H.~Y.} \bibnamefont{Li}},
  \bibinfo{author}{\bibfnamefont{D.~J.~J.} \bibnamefont{Farnell}},
  \bibnamefont{and} \bibinfo{author}{\bibfnamefont{C.~E.}
  \bibnamefont{Campbell}}, \bibinfo{journal}{J. Phys.: Condens. Matter}
  \textbf{\bibinfo{volume}{25}}, \bibinfo{pages}{306002}
  (\bibinfo{year}{2013}).

\bibitem[{\citenamefont{Gong et~al.}(2013)\citenamefont{Gong, Sheng, Motrunich,
  and Fisher}}]{Fisher_2013}
\bibinfo{author}{\bibfnamefont{S.-S.} \bibnamefont{Gong}},
  \bibinfo{author}{\bibfnamefont{D.}~\bibnamefont{Sheng}},
  \bibinfo{author}{\bibfnamefont{O.~I.} \bibnamefont{Motrunich}},
  \bibnamefont{and} \bibinfo{author}{\bibfnamefont{M.~P.}
  \bibnamefont{Fisher}}, \bibinfo{journal}{Phys. Rev. B}
  \textbf{\bibinfo{volume}{88}}, \bibinfo{pages}{165138}
  (\bibinfo{year}{2013}).

\bibitem[{\citenamefont{Ganesh et~al.}(2013)\citenamefont{Ganesh, van~den
  Brink, and Nishimoto}}]{Ganesh_PRL_2013}
\bibinfo{author}{\bibfnamefont{R.}~\bibnamefont{Ganesh}},
  \bibinfo{author}{\bibfnamefont{J.}~\bibnamefont{van~den Brink}},
  \bibnamefont{and}
  \bibinfo{author}{\bibfnamefont{S.}~\bibnamefont{Nishimoto}},
  \bibinfo{journal}{Phys. Rev. Lett.} \textbf{\bibinfo{volume}{110}},
  \bibinfo{pages}{127203} (\bibinfo{year}{2013}).

\bibitem[{\citenamefont{Zhu et~al.}(2013)\citenamefont{Zhu, Huse, and
  White}}]{Zhu_PRL_2013}
\bibinfo{author}{\bibfnamefont{Z.}~\bibnamefont{Zhu}},
  \bibinfo{author}{\bibfnamefont{D.~A.} \bibnamefont{Huse}}, \bibnamefont{and}
  \bibinfo{author}{\bibfnamefont{S.~R.} \bibnamefont{White}},
  \bibinfo{journal}{Phys. Rev. Lett.} \textbf{\bibinfo{volume}{110}},
  \bibinfo{pages}{127205} (\bibinfo{year}{2013}).

\bibitem[{\citenamefont{Zhang and Lamas}(2013)}]{Zhang_PRB_2013}
\bibinfo{author}{\bibfnamefont{H.}~\bibnamefont{Zhang}} \bibnamefont{and}
  \bibinfo{author}{\bibfnamefont{C.}~\bibnamefont{Lamas}},
  \bibinfo{journal}{Phys. Rev. B} \textbf{\bibinfo{volume}{87}},
  \bibinfo{pages}{024415} (\bibinfo{year}{2013}).


\bibitem[{\citenamefont{Smirnova et~al.}(2009)\citenamefont{Smirnova, Azuma,
  Kumada, Kusano, Matsuda, Shimakawa, Takei, Yonesaki, and
  Kinomura}}]{smirnova2009synthesis}
\bibinfo{author}{\bibfnamefont{O.}~\bibnamefont{Smirnova}},
  \bibinfo{author}{\bibfnamefont{M.}~\bibnamefont{Azuma}},
  \bibinfo{author}{\bibfnamefont{N.}~\bibnamefont{Kumada}},
  \bibinfo{author}{\bibfnamefont{Y.}~\bibnamefont{Kusano}},
  \bibinfo{author}{\bibfnamefont{M.}~\bibnamefont{Matsuda}},
  \bibinfo{author}{\bibfnamefont{Y.}~\bibnamefont{Shimakawa}},
  \bibinfo{author}{\bibfnamefont{T.}~\bibnamefont{Takei}},
  \bibinfo{author}{\bibfnamefont{Y.}~\bibnamefont{Yonesaki}}, \bibnamefont{and}
  \bibinfo{author}{\bibfnamefont{N.}~\bibnamefont{Kinomura}},
  \bibinfo{journal}{Journal of the American Chemical Society}
  \textbf{\bibinfo{volume}{131}}, \bibinfo{pages}{8313} (\bibinfo{year}{2009}).


\bibitem[{\citenamefont{Kandpal and van~den
  Brink}(2011)}]{kandpal2011calculation}
\bibinfo{author}{\bibfnamefont{H.~C.} \bibnamefont{Kandpal}} \bibnamefont{and}
  \bibinfo{author}{\bibfnamefont{J.}~\bibnamefont{van~den Brink}},
  \bibinfo{journal}{Physical Review B} \textbf{\bibinfo{volume}{83}},
  \bibinfo{pages}{140412} (\bibinfo{year}{2011}).

\bibitem[{\citenamefont{Matsuda et~al.}(2010)\citenamefont{Matsuda, Azuma,
  Tokunaga, Shimakawa, and Kumada}}]{expnew2}
\bibinfo{author}{\bibfnamefont{M.}~\bibnamefont{Matsuda}},
  \bibinfo{author}{\bibfnamefont{M.}~\bibnamefont{Azuma}},
  \bibinfo{author}{\bibfnamefont{M.}~\bibnamefont{Tokunaga}},
  \bibinfo{author}{\bibfnamefont{Y.}~\bibnamefont{Shimakawa}},
  \bibnamefont{and} \bibinfo{author}{\bibfnamefont{N.}~\bibnamefont{Kumada}},
  \bibinfo{journal}{Phys. Rev. Lett.} \textbf{\bibinfo{volume}{105}},
  \bibinfo{pages}{187201} (\bibinfo{year}{2010}).

\bibitem[{\citenamefont{Ganesh et~al.}(2011{\natexlab{b}})\citenamefont{Ganesh,
  Isakov, and Paramekanti}}]{Ganesh_QMC}
\bibinfo{author}{\bibfnamefont{R.}~\bibnamefont{Ganesh}},
  \bibinfo{author}{\bibfnamefont{S.~V.} \bibnamefont{Isakov}},
  \bibnamefont{and}
  \bibinfo{author}{\bibfnamefont{A.}~\bibnamefont{Paramekanti}},
  \bibinfo{journal}{Phys. Rev. B} \textbf{\bibinfo{volume}{84}},
  \bibinfo{pages}{214412} (\bibinfo{year}{2011}{\natexlab{b}}).

\bibitem[{\citenamefont{Oitmaa and Singh}(2012)}]{Oitmaa_2012}
\bibinfo{author}{\bibfnamefont{J.}~\bibnamefont{Oitmaa}} \bibnamefont{and}
  \bibinfo{author}{\bibfnamefont{R.}~\bibnamefont{Singh}},
  \bibinfo{journal}{Phys. Rev. B} \textbf{\bibinfo{volume}{85}},
  \bibinfo{pages}{014428} (\bibinfo{year}{2012}).

\bibitem[{\citenamefont{Zhang et~al.}(2014)\citenamefont{Zhang, Arlego, and
  Lamas}}]{Zhang2014}
\bibinfo{author}{\bibfnamefont{H.}~\bibnamefont{Zhang}},
  \bibinfo{author}{\bibfnamefont{M.}~\bibnamefont{Arlego}}, \bibnamefont{and}
  \bibinfo{author}{\bibfnamefont{C.~A.} \bibnamefont{Lamas}},
  \bibinfo{journal}{Phys. Rev. B} \textbf{\bibinfo{volume}{89}},
  \bibinfo{pages}{024403} (\bibinfo{year}{2014}).

\bibitem[{\citenamefont{Arlego et~al.}(2014)\citenamefont{Arlego, Lamas, and
  Zhang}}]{Arlego201415}
\bibinfo{author}{\bibfnamefont{M.}~\bibnamefont{Arlego}},
  \bibinfo{author}{\bibfnamefont{C.~A.} \bibnamefont{Lamas}}, \bibnamefont{and}
  \bibinfo{author}{\bibfnamefont{H.}~\bibnamefont{Zhang}}, \bibinfo{journal}{J.
  Phys.: Conf. Ser.} \textbf{\bibinfo{volume}{568}}, \bibinfo{pages}{042019}
  (\bibinfo{year}{2014}).

\bibitem[{\citenamefont{Zhang et~al.}(2016)\citenamefont{Zhang, Lamas, Arlego,
  and Brenig}}]{Brenig2016}
\bibinfo{author}{\bibfnamefont{H.}~\bibnamefont{Zhang}},
  \bibinfo{author}{\bibfnamefont{C.~A.} \bibnamefont{Lamas}},
  \bibinfo{author}{\bibfnamefont{M.}~\bibnamefont{Arlego}}, \bibnamefont{and}
  \bibinfo{author}{\bibfnamefont{W.}~\bibnamefont{Brenig}},
  \bibinfo{journal}{Phys. Rev. B} \textbf{\bibinfo{volume}{93}},
  \bibinfo{pages}{235150} (\bibinfo{year}{2016}).

\bibitem[{\citenamefont{Bishop and Li}(2017)}]{bishop2017frustrated}
\bibinfo{author}{\bibfnamefont{R.~F.} \bibnamefont{Bishop}} \bibnamefont{and}
  \bibinfo{author}{\bibfnamefont{P.~H.~Y.} \bibnamefont{Li}},
  \bibinfo{journal}{Phys. Rev. B} \textbf{\bibinfo{volume}{95}},
  \bibinfo{pages}{134414} (\bibinfo{year}{2017}).

\bibitem[{\citenamefont{Krokhmalskii et~al.}(2017)\citenamefont{Krokhmalskii,
  Baliha, Derzhko, Schulenburg, and Richter}}]{Richter2017}
\bibinfo{author}{\bibfnamefont{T.}~\bibnamefont{Krokhmalskii}},
  \bibinfo{author}{\bibfnamefont{V.}~\bibnamefont{Baliha}},
  \bibinfo{author}{\bibfnamefont{O.}~\bibnamefont{Derzhko}},
  \bibinfo{author}{\bibfnamefont{J.}~\bibnamefont{Schulenburg}},
  \bibnamefont{and} \bibinfo{author}{\bibfnamefont{J.}~\bibnamefont{Richter}},
  \bibinfo{journal}{Phys. Rev. B} \textbf{\bibinfo{volume}{95}},
  \bibinfo{pages}{094419} (\bibinfo{year}{2017}).

\bibitem[{\citenamefont{Bose}(1992)}]{Bose1992}
\bibinfo{author}{\bibfnamefont{I.}~\bibnamefont{Bose}}, \bibinfo{journal}{Phys.
  Rev. B} \textbf{\bibinfo{volume}{45}}, \bibinfo{pages}{13072}
  (\bibinfo{year}{1992}).

\bibitem[{\citenamefont{Bose and Gayen}(1993)}]{Bose1993a}
\bibinfo{author}{\bibfnamefont{I.}~\bibnamefont{Bose}} \bibnamefont{and}
  \bibinfo{author}{\bibfnamefont{S.}~\bibnamefont{Gayen}},
  \bibinfo{journal}{Phys. Rev. B} \textbf{\bibinfo{volume}{48}},
  \bibinfo{pages}{10653(R)} (\bibinfo{year}{1993}).

\bibitem[{\citenamefont{Honecker et~al.}(2000)\citenamefont{Honecker, Mila, and
  Troyer}}]{Honecker2000a}
\bibinfo{author}{\bibfnamefont{A.}~\bibnamefont{Honecker}},
  \bibinfo{author}{\bibfnamefont{F.}~\bibnamefont{Mila}}, \bibnamefont{and}
  \bibinfo{author}{\bibfnamefont{M.}~\bibnamefont{Troyer}},
  \bibinfo{journal}{Eur. Phys. J. B} \textbf{\bibinfo{volume}{15}},
  \bibinfo{pages}{227} (\bibinfo{year}{2000}).

\bibitem[{\citenamefont{Brenig and Becker}(2001)}]{Brenig2001a}
\bibinfo{author}{\bibfnamefont{W.}~\bibnamefont{Brenig}} \bibnamefont{and}
  \bibinfo{author}{\bibfnamefont{K.~W.} \bibnamefont{Becker}},
  \bibinfo{journal}{Phys. Rev. B} \textbf{\bibinfo{volume}{64}},
  \bibinfo{pages}{214413} (\bibinfo{year}{2001}).

\bibitem[{\citenamefont{Arlego and Brenig}(2006)}]{Arlego2006a}
\bibinfo{author}{\bibfnamefont{M.}~\bibnamefont{Arlego}} \bibnamefont{and}
  \bibinfo{author}{\bibfnamefont{W.}~\bibnamefont{Brenig}},
  \bibinfo{journal}{Eur. Phys. J. B} \textbf{\bibinfo{volume}{53}},
  \bibinfo{pages}{193} (\bibinfo{year}{2006}).

\bibitem[{\citenamefont{Lamas and Matera}(2015)}]{Lamas2015b}
\bibinfo{author}{\bibfnamefont{C.~A.} \bibnamefont{Lamas}} \bibnamefont{and}
  \bibinfo{author}{\bibfnamefont{J.~M.} \bibnamefont{Matera}},
  \bibinfo{journal}{Phys. Rev. B} \textbf{\bibinfo{volume}{92}},
  \bibinfo{pages}{115111} (\bibinfo{year}{2015}).


  \bibitem{matera-lamas2}
  J.M. Matera, C.A. Lamas, J. Phys.: Condens. Matter 26 (2014) 326004

\bibitem[{\citenamefont{Sachdev and Bhatt}(1990)}]{Sachdev1990}
\bibinfo{author}{\bibfnamefont{S.}~\bibnamefont{Sachdev}} \bibnamefont{and}
  \bibinfo{author}{\bibfnamefont{R.~N.} \bibnamefont{Bhatt}},
  \bibinfo{journal}{Phys. Rev. B} \textbf{\bibinfo{volume}{41}},
  \bibinfo{pages}{9323} (\bibinfo{year}{1990}).

\bibitem[{\citenamefont{Knetter and Uhrig}(2000)}]{SE-CUT}
\bibinfo{author}{\bibfnamefont{C.}~\bibnamefont{Knetter}} \bibnamefont{and}
  \bibinfo{author}{\bibfnamefont{G.~S.} \bibnamefont{Uhrig}},
  \bibinfo{journal}{The European Physical Journal B-Condensed Matter and
  Complex Systems} \textbf{\bibinfo{volume}{13}}, \bibinfo{pages}{209}
  (\bibinfo{year}{2000}).

\bibitem[{\citenamefont{Auerbach and Arovas}(1988)}]{Auerbach}
\bibinfo{author}{\bibfnamefont{A.}~\bibnamefont{Auerbach}} \bibnamefont{and}
  \bibinfo{author}{\bibfnamefont{D.~P.} \bibnamefont{Arovas}},
  \bibinfo{journal}{Phys. Rev. Lett.} \textbf{\bibinfo{volume}{61}},
  \bibinfo{pages}{617} (\bibinfo{year}{1988}).

\bibitem[{\citenamefont{Auerbach}(1994)}]{Auerbach:1994}
\bibinfo{author}{\bibfnamefont{A.}~\bibnamefont{Auerbach}},
  \emph{\bibinfo{title}{Interacting Electrons and Quantum Magnetism}}
  (\bibinfo{publisher}{Springer-Verlag}, \bibinfo{address}{New York},
  \bibinfo{year}{1994}).

\bibitem[{\citenamefont{Auerbach and Arovas}(2011)}]{Auerbach:2011}
\bibinfo{author}{\bibfnamefont{A.}~\bibnamefont{Auerbach}} \bibnamefont{and}
  \bibinfo{author}{\bibfnamefont{D.~P.} \bibnamefont{Arovas}}, in
  \emph{\bibinfo{booktitle}{Introduction to Frustrated Magnetism}}
  (\bibinfo{publisher}{Springer}, \bibinfo{year}{2011}).

\bibitem[{\citenamefont{Ceccatto et~al.}(1993)\citenamefont{Ceccatto, Gazza,
  and Trumper}}]{Trumper1}
\bibinfo{author}{\bibfnamefont{H.}~\bibnamefont{Ceccatto}},
  \bibinfo{author}{\bibfnamefont{C.}~\bibnamefont{Gazza}}, \bibnamefont{and}
  \bibinfo{author}{\bibfnamefont{A.}~\bibnamefont{Trumper}},
  \bibinfo{journal}{Phys. Rev. B} \textbf{\bibinfo{volume}{47}},
  \bibinfo{pages}{12329} (\bibinfo{year}{1993}).

\bibitem[{\citenamefont{Trumper et~al.}(1997)\citenamefont{Trumper, Manuel,
  Gazza, and Ceccatto}}]{Trumper2}
\bibinfo{author}{\bibfnamefont{A.}~\bibnamefont{Trumper}},
  \bibinfo{author}{\bibfnamefont{L.}~\bibnamefont{Manuel}},
  \bibinfo{author}{\bibfnamefont{C.}~\bibnamefont{Gazza}}, \bibnamefont{and}
  \bibinfo{author}{\bibfnamefont{H.}~\bibnamefont{Ceccatto}},
  \bibinfo{journal}{Phys. Rev. Lett.} \textbf{\bibinfo{volume}{78}},
  \bibinfo{pages}{2216} (\bibinfo{year}{1997}).

\bibitem[{\citenamefont{Flint and Coleman}(2009)}]{Coleman}
\bibinfo{author}{\bibfnamefont{R.}~\bibnamefont{Flint}} \bibnamefont{and}
  \bibinfo{author}{\bibfnamefont{P.}~\bibnamefont{Coleman}},
  \bibinfo{journal}{Physical Review B} \textbf{\bibinfo{volume}{79}},
  \bibinfo{pages}{014424} (\bibinfo{year}{2009}).

\bibitem[{\citenamefont{Mezio et~al.}(2011)\citenamefont{Mezio, Sposetti,
  Manuel, and Trumper}}]{Mezio}
\bibinfo{author}{\bibfnamefont{A.}~\bibnamefont{Mezio}},
  \bibinfo{author}{\bibfnamefont{C.}~\bibnamefont{Sposetti}},
  \bibinfo{author}{\bibfnamefont{L.}~\bibnamefont{Manuel}}, \bibnamefont{and}
  \bibinfo{author}{\bibfnamefont{A.}~\bibnamefont{Trumper}},
  \bibinfo{journal}{Eur. Phys. Lett.} \textbf{\bibinfo{volume}{94}},
  \bibinfo{pages}{47001} (\bibinfo{year}{2011}).

\bibitem[{\citenamefont{Messio et~al.}(2012)\citenamefont{Messio, Bernu, and
  Lhuillier}}]{Messio}
\bibinfo{author}{\bibfnamefont{L.}~\bibnamefont{Messio}},
  \bibinfo{author}{\bibfnamefont{B.}~\bibnamefont{Bernu}}, \bibnamefont{and}
  \bibinfo{author}{\bibfnamefont{C.}~\bibnamefont{Lhuillier}},
  \bibinfo{journal}{Phys. Rev. Lett.} \textbf{\bibinfo{volume}{108}},
  \bibinfo{pages}{207204} (\bibinfo{year}{2012}).

\bibitem[{\citenamefont{Messio et~al.}(2013)\citenamefont{Messio, Lhuillier,
  and Misguich}}]{Messio_2013}
\bibinfo{author}{\bibfnamefont{L.}~\bibnamefont{Messio}},
  \bibinfo{author}{\bibfnamefont{C.}~\bibnamefont{Lhuillier}},
  \bibnamefont{and} \bibinfo{author}{\bibfnamefont{G.}~\bibnamefont{Misguich}},
  \bibinfo{journal}{Phys. Rev. B} \textbf{\bibinfo{volume}{87}},
  \bibinfo{pages}{125127} (\bibinfo{year}{2013}).

\bibitem{note:corr}
For the use of correlations and staggered magnetization to clasify magnetic
phases see for example Phys. Rev. B {\bf 83}, 094506 (2011).

\bibitem[{\citenamefont{Oitmaa et~al.}(2006)\citenamefont{Oitmaa, Hamer, and
  Zheng}}]{Book-SE}
\bibinfo{author}{\bibfnamefont{J.}~\bibnamefont{Oitmaa}},
  \bibinfo{author}{\bibfnamefont{C.}~\bibnamefont{Hamer}}, \bibnamefont{and}
  \bibinfo{author}{\bibfnamefont{W.}~\bibnamefont{Zheng}},
  \emph{\bibinfo{title}{Series expansion methods for strongly interacting
  lattice models}} (\bibinfo{publisher}{Cambridge University Press},
  \bibinfo{year}{2006}).


\bibitem[{\citenamefont{Chubukov}(1989)}]{Chubukov1989}
\bibinfo{author}{\bibfnamefont{A.~V.} \bibnamefont{Chubukov}},
  \bibinfo{journal}{JETP Lett.} \textbf{\bibinfo{volume}{49}},
  \bibinfo{pages}{129} (\bibinfo{year}{1989}).

\bibitem[{\citenamefont{Chubukov and Jolicoeur}(1991)}]{Chubukov1991}
\bibinfo{author}{\bibfnamefont{A.~V.} \bibnamefont{Chubukov}} \bibnamefont{and}
  \bibinfo{author}{\bibfnamefont{T.}~\bibnamefont{Jolicoeur}},
  \bibinfo{journal}{Phys. Rev. B} \textbf{\bibinfo{volume}{44}},
  \bibinfo{pages}{12050(R)} (\bibinfo{year}{1991}).

\bibitem{Kotov1998}
V. N. Kotov, O. Sushkov, Zheng Weihong, and J. Oitmaa,
Phys. Rev. Lett. {\bf 80}, 5970 (1998).

\bibitem{Matsumoto2002}
M. Matsumoto, B. Normand, T. M. Rice, and M. Sigrist,
Phys. Rev. Lett. {\bf 89}, 077203 (2002).


\bibitem[{\citenamefont{Wegner}(1994)}]{Wegner-1994a}
\bibinfo{author}{\bibfnamefont{F.}~\bibnamefont{Wegner}},
  \bibinfo{journal}{Ann. Phys.} \textbf{\bibinfo{volume}{506}},
  \bibinfo{pages}{77} (\bibinfo{year}{1994}).

\bibitem[{\citenamefont{Arlego and Brenig}(2011)}]{Arlego2011}
\bibinfo{author}{\bibfnamefont{M.}~\bibnamefont{Arlego}} \bibnamefont{and}
  \bibinfo{author}{\bibfnamefont{W.}~\bibnamefont{Brenig}},
  \bibinfo{journal}{Phys. Rev. B} \textbf{\bibinfo{volume}{84}},
  \bibinfo{pages}{134426} (\bibinfo{year}{2011}).

\bibitem[{\citenamefont{Arlego and Brenig}(2008)}]{Arlego2008a}
\bibinfo{author}{\bibfnamefont{M.}~\bibnamefont{Arlego}} \bibnamefont{and}
  \bibinfo{author}{\bibfnamefont{W.}~\bibnamefont{Brenig}},
  \bibinfo{journal}{Phys. Rev. B} \textbf{\bibinfo{volume}{78}},
  \bibinfo{pages}{224415} (\bibinfo{year}{2008}).

\bibitem[{\citenamefont{Arlego and Brenig}(2007)}]{Arlego2007a}
\bibinfo{author}{\bibfnamefont{M.}~\bibnamefont{Arlego}} \bibnamefont{and}
  \bibinfo{author}{\bibfnamefont{W.}~\bibnamefont{Brenig}},
  \bibinfo{journal}{Phys. Rev. B} \textbf{\bibinfo{volume}{75}},
  \bibinfo{pages}{024409} (\bibinfo{year}{2007}).

\bibitem{SuppMat}
The explicit form of these expressions is impractical to be presented here in written
form, but they are available in the supplemental material.

\end{thebibliography}
\end{document}